\documentclass[11pt, a4paper]{article}
\usepackage{amsmath}

\newtheorem{theorem}{Theorem}

\newcommand{\benumerate}{\begin{enumerate}}
\newcommand{\eenumerate}{\end{enumerate}}

\newcommand{\bitemize}{\begin{itemize}}

\newcommand{\eitemize}{\end{itemize}}

\begin{document}

\title{On the central quadric ansatz:  integrable models and Painlev\'e reductions}
\author{E.V. Ferapontov, B. Huard and A. Zhang}
    \date{}
    \maketitle
    \vspace{-7mm}
\begin{center}
Department of Mathematical Sciences \\ Loughborough University \\
Loughborough, Leicestershire LE11 3TU \\ United Kingdom \\[2ex]
e-mails: \\  { \texttt{E.V.Ferapontov@lboro.ac.uk}}\\

{\texttt{B.Huard@lboro.ac.uk}}\\

{\texttt{A.Zhang-10@student.lboro.ac.uk}}\\

\end{center}

\bigskip

\begin{abstract}

It was observed by Tod  \cite{Tod} and later by Dunajski and Tod \cite{Dunajski} that the Boyer-Finley (BF) and the dispersionless Kadomtsev-Petviashvili (dKP) equations possess solutions whose level surfaces are central quadrics in the space of independent variables (the so-called central quadric ansatz). It was demonstrated that generic  solutions of this type are described by  Painlev\'e equations $P_{III}$ and $P_{II}$, respectively.
The aim of our paper is threefold:

\noindent -- Based on the method of hydrodynamic reductions, we classify integrable models possessing the central quadric ansatz. This leads to the five canonical forms (including BF and dKP). 

\noindent -- Applying the central quadric ansatz to each of the five  canonical forms, we obtain all Painlev\'e equations $P_{I} - P_{VI}$, with $P_{VI}$ corresponding to the generic case of our classification.

\noindent --  We argue that solutions coming from the central quadric ansatz constitute a subclass of two-phase solutions provided by the method of hydrodynamic reductions.

\bigskip

\noindent MSC: 34M55, 35C05, 35L10, 35Q51, 35Q75, 37K10.

\bigskip

Keywords: multidimensional  integrability, hydrodynamic reductions,
 central quadric ansatz, Painlev\'e equations.
\end{abstract}

\newpage

\section{Introduction}

This paper grew from an attempt to understand the construction of \cite{Tod, Dunajski} reducing the Boyer-Finley (BF) and the dispersionless Kadomtsev-Petviashvili (dKP) equations to Painlev\'e trans\-cendents via the so-called `central quadric ansatz'. This procedure applies to PDEs of the form
\begin{equation}
\label{eq:original}
(a(u))_{xx} + (b(u))_{yy} + (c(u))_{tt} + 2 (p(u))_{xy} + 2 (q(u))_{xt} + 2 (r(u))_{yt} = 0,
\end{equation}
and consists of seeking solutions $u(x, y, t)$ in implicit form, 
\begin{equation}
(x, y, t) M(u) (x, y, t)^T=1,
\label{central}
\end{equation}
where $M(u)$ is a $3\times 3$ symmetric matrix of $u$. The level surfaces of such solutions, $u=const$, are central quadrics in the space of independent variables $x, y, t$. Remarkably, for both   BF and dKP equations considered in \cite{Tod, Dunajski}, differential equations for the matrix $M(u)$  reduce to Painlev\'e trans\-cendents. Although the occurrence of Painlev\'e transcendents via similarity reductions of soliton equations is a well-known fact, their appearance in the context of multidimensional {\it dispersionless} integrable PDEs is, to the best of our knowledge,  an entirely new phenomenon. In this paper we address the following questions:

\begin{itemize}
\item Classify  integrable PDEs of the form (\ref{eq:original}). Our approach to this problem is based on the method of hydrodynamic reductions which, as demonstrated in \cite{Fer}, provides  an efficient criterion for the classification of multidimensional dispersionless integrable equations. In particular, both BF and dKP equations are known to possess an infinity of hydrodynamic reductions.
The classification is to be performed modulo (complex) linear changes of the independent variables $x, y, t$, as well as transformations $u\to \varphi(u)$, which constitute the equivalence group of our problem.

\item Describe Painlev\'e reductions of integrable PDEs resulting from the above classification by applying the central quadric ansatz.

\item Relate the central quadric ansatz to the method of hydrodynamic reductions.

\end{itemize}

\noindent To formulate our first  result we introduce the symmetric matrix
$$
V(u)=\left(\begin{array}{ccc}
a'&p'&q'\\
p'&b'&r'\\
q'&r'&c'
\end{array}
\right),
$$
where prime denotes differentiation by $u$.

\begin{theorem} A PDE of the form (\ref{eq:original}) is integrable by the method of hydrodynamic reductions if and only if the matrix $V(u)$ satisfies the constraint
\begin{equation}
V''=(\ln \det V)' V'+kV,
\label{b}
\end{equation}
for some scalar function k. 
Modulo equivalence transformations, this leads to the five canonical forms of nonlinear integrable models possessing the central quadric ansatz:
$$
u_{xx}+u_{yy}-g(u)_{yy}-g(u)_{tt}=0,
$$
$$
u_{xx}+u_{yy}+(e^u)_{tt}=0,
$$
$$
(e^u-u)_{xx}+2u_{xy}+(e^u)_{tt}=0,
$$
$$
u_{xt}-(uu_x)_x-u_{yy}=0,
$$
$$
(u^2)_{xy}+u_{yy}+2u_{xt}=0,
$$
here $g(u)=\ln(1-e^u)$. Examples 2 and 4 are the familiar BF and dKP equations.

\end{theorem}

We point out that the constraint (\ref{b}), which  implies  $V''\in span \{V, V'\}$,  means that the `curve' $V(u)$ lies in a two-dimensional linear subspace of the space of $3\times 3$ symmetric matrices.  The classification of normal forms of such linear subspaces  leads to the five canonical forms of Theorem 1.  Equations listed in Theorem 1 are not new: in different guises, they have appeared in the classification of multidimensional integrable systems (see Sect. 2 for the proof  of Theorem 1 and further discussion).

\medskip

In Sect. 3 we apply the central quadric ansatz (\ref{central}) to  the five canonical forms  of Theorem 1. Differentiating  (\ref{central}) implicitly and substituting into  (\ref{eq:original}) one obtains  equations for the matrix $M(u)$ which can be most naturally written  in terms of the inverse matrix $N=-M^{-1}$:
\begin{equation}
gN'=V ~~~ {\rm where} ~~~ g^2 \det N=\xi=const,
\label{N}
\end{equation}
see   \cite{Tod, Dunajski}. 
For matrices  $V$ associated with the five canonical forms of Theorem 1,  {\it generic} solutions of  equations (\ref{N})  reduce to the following Painlev\'e transcendents:

\medskip

\noindent {\it Case 1:} General case of $P_{VI}$;

\noindent {\it Case 2:} Special case of   $P_{V}$ reducible to $P_{III}$;

\noindent {\it Case 3:} General case of $P_{V}$;

\noindent {\it Case 4:} General case of $P_{II}$ with a reduction to $P_I$;

\noindent {\it Case 5:} General case of $P_{IV}$.

 \medskip
 
 It is quite remarkable that the whole variety of Painlev\'e equations is contained in the pair of  matrix equations (\ref{b}), (\ref{N}): setting $\xi=1$ and eliminating the scalar $g$ we can rewrite them 
 in the equivalent form,
 $$
 N'=V \sqrt {\det N}, ~~~ (V'/ \det V)' =\kappa V.
$$

In Sect. 4 we  show that, at least for the dKP equation, solutions coming from the central quadric ansatz constitute a particular subclass of two-phase solutions provided by the method of hydrodynamic reductions.

\section{Classification of integrable PDEs of the form (\ref{eq:original}): proof of Theorem 1}

Following the method of hydrodynamic reductions \cite{GibTsa96, GibTsa99,Fer}, we seek multi-phase solutions of   equation (\ref{eq:original}) in the form 
\begin{equation}
\label{eq:hydrodynamic-u}
u=u(R^1, \ldots, R^N),
\end{equation}
where the phases $R^i(x,y, t)$  are required  to satisfy a pair of compatible  systems of hydrodynamic type,
\begin{equation}
\label{eq:hydrodynamic}
R^i_t = \lambda^i(R) R^i_x, \quad R^i_y = \mu^i(R) R^i_x,
\end{equation}
$R = (R^1, \ldots, R^N)$.  We say that equations (\ref{eq:hydrodynamic}) provide an $N$-component  hydrodynamic reduction of the equation  (\ref{eq:original}). Note that the number of phases, $N$, is allowed to be arbitrary. The compatibility of equations (\ref{eq:hydrodynamic}) requires the following conditions \cite{Tsarev}, 
\begin{equation}
\label{eq:Gibbons-Tsarev}
\frac{\lambda^i_j}{\lambda^j - \lambda^i} = \frac{\mu^i_j }{\mu^j - \mu^i}, 
\end{equation}
 $ i\neq j$,  here $  \lambda^i_j = \partial_{R^j} \lambda^i, \ \mu^i_j =\partial_{R^j} \mu^i$. It was observed in \cite{Fer} that the requirement of the existence of $N$-phase solutions parametrized by $N$ arbitrary functions of one variable imposes strong constraints on the original system (\ref{eq:original}), and provides an efficient approach to the classification of multidimensional dispersionless integrable  systems. 

 In the present context the integrability conditions can be derived as follows. Upon substitution of (\ref{eq:hydrodynamic-u}) and (\ref{eq:hydrodynamic}) into equation (\ref{eq:original}) we obtain that the characteristic speeds $\lambda^i$ and $ \mu^i$ obey the dispersion relation,
\begin{equation}
\label{eq:dispersion}
\Delta^i=a' + b' {\mu^i}^2 + c' {\lambda^i}^2 + 2 \left(p' \mu^i  + q' \lambda^i + r' \lambda^i \mu^i \right) = 0.
\end{equation}
Differentiating this relation with respect to $R^j$,  $ i\neq j$,  and taking into account the equations (\ref{eq:Gibbons-Tsarev}) and (\ref{eq:dispersion}), we obtain 
\begin{equation}
\label{eq:l-derivatives}
\begin{split}
\lambda^i_j = \frac{1}{2} (\lambda^i - \lambda^j) \frac{\Dot{\Delta}^{i}}{\Delta^{ij}} u_j, \quad 
\mu^i_j =  \frac{1}{2} (\mu^i - \mu^j) \frac{\Dot{\Delta}^{i}}{\Delta^{ij}} u_j,
\end{split}
\end{equation}
where we introduced the  notation
\begin{equation*}
\begin{split}
&\Dot{\Delta}^i = a'' + b'' {\mu^i}^2 + c'' {\lambda^i}^2 + 2 \left(p'' \mu^i  + q'' \lambda^i + r'' \lambda^i \mu^i \right), \\
&\Delta^{ij} = a' + b' {\mu^i} \mu^j + c' {\lambda^i} \lambda^j + p' (\mu^i+\mu^j)  + q'  (\lambda^i + \lambda^j) + r' (\lambda^i \mu^i +\lambda^j \mu^i ). 
\end{split}
\end{equation*}
Moreover, we obtain the following expressions for the second order derivatives of $u$, 
\begin{equation}
u_{ij} = - \frac{\dot{\Delta}^{ij}}{ \Delta^{ij}}u_i u_j ,
\label{u}
\end{equation}
where
$$
\Dot \Delta^{ij} = a'' + b'' {\mu^i} \mu^j + c'' {\lambda^i} \lambda^j + p'' (\mu^i+\mu^j)  + q''  (\lambda^i + \lambda^j) + r'' (\lambda^i \mu^i +\lambda^j \mu^i ). 
$$
Equations (\ref{eq:l-derivatives}) and (\ref{u}) constitute the so-called generalised Gibbons-Tsarev system governing multi-phase solutions (hydrodynamic reductions)  of equation (\ref{eq:original}).
The  conditions of compatibility of Eqs  (\ref{eq:l-derivatives})-(\ref{u}), $(\lambda^i_j)_k=(\lambda^i_k)_j$, $(\mu^i_j)_k=(\mu^i_k)_j$   and $(u_{ij})_k=(u_{ik})_j$, 
lead to polynomial expressions in the characteristic speeds $\lambda^i, \lambda^j, \lambda^k$ and $\mu^i, \mu^j, \mu^k$ which are required to  vanish modulo  the dispersion relation (\ref{eq:dispersion}). This leads to the integrability conditions (\ref{b}). 

\bigskip

To solve Eq. (\ref{b}) let us view $V(u)$ as a curve in the space of $3\times 3$ symmetric matrices. Since the acceleration vector $V''$ is a linear combination of the position vector $V$ and the velocity vector $V'$, this curve must belong to a fixed two-dimensional linear subspace (the case when this subspace is one-dimensional corresponds to linear equations: in this case the central quadric ansatz was analysed in \cite{Dunajski1}).  The classification of two-dimensional linear subspaces of  the space of $3\times 3$ symmetric matrices leads to the five canonical forms which are presented below in the format $\it span\{A, B\}$ for some particular choices of symmetric matrices $A$ and $B$. Recall that pairs of symmetric matrices, $\tilde A$ and $\tilde B$,  are classified by  Jordan normal forms of the operator $\tilde B\tilde A^{-1}$, further subcases correspond to coincidences among eigenvalues of this operator. Choosing a suitable basis $A, B$ in the subspace $span\{\tilde A, \tilde B\}$ leads to the five canonical forms mentioned above.  In each of these cases one can set $V=A+f(u)B$ (to be precise, $V=g(u)A+f(u)B$, however, $g(u)$ can be set equal to $1$ via an equivalence transformation $u\to \varphi(u)$). The substitution into the integrability condition (\ref{b}) leads to  simple ODEs for $f(u)$ which can be readily solved. 
\medskip

\noindent {\bf Case 1.}
$$
A=\left(\begin{array}{ccc}
1&0&0\\
0&1&0\\
0&0&0
\end{array}
\right), ~~~ 
B=\left(\begin{array}{ccc}
0&0&0\\
0&1&0\\
0&0&1
\end{array}
\right).
$$
The substitution of $V=A+f(u)B$ into the integrability condition (\ref{b}) leads to $k=0$ and $f''=f'^2\left(\frac{1}{f}+\frac{1}{f+1}\right)$, so that without any loss of generality one can take $f(u)=e^u/(1-e^u)$. This leads to the first equation of Theorem 1. Setting   $u=v_{yy}+v_{tt}$ we obtain the equation $e^{v_{xx}+v_{yy}}+e^{v_{tt}+v_{yy}}=1$ or, equivalently,  $e^{v_{xx}}+e^{v_{tt}}=e^{-v_{yy}}$, which first appeared in \cite{Fer1} in the classification of integrable PDEs of the form $F(v_{xx}, v_{yy}, v_{tt})=0$. It can be viewed as an integrable generalisation of the BF equation.

\medskip

\noindent {\bf Case 2.}
$$
A=\left(\begin{array}{ccc}
1&0&0\\
0&1&0\\
0&0&0
\end{array}
\right), ~~~ 
B=\left(\begin{array}{ccc}
0&0&0\\
0&0&0\\
0&0&1
\end{array}
\right).
$$
The substitution of $V=A+f(u)B$ into the integrability condition (\ref{b}) leads to $k=0$ and $f''=f'^2/f$, so that without any loss of generality one can take $f(u)=e^u$. This leads to the BF equation.

\medskip

\noindent {\bf Case 3.}
$$
A=\left(\begin{array}{ccc}
0&1&0\\
1&0&0\\
0&0&1
\end{array}
\right), ~~~ 
B=\left(\begin{array}{ccc}
1&0&0\\
0&0&0\\
0&0&1
\end{array}
\right).
$$
The substitution of $V=A+f(u)B$ into the integrability condition (\ref{b}) leads to $k=0$ and $f''=f'^2/(f+1)$, so that without any loss of generality one can take $f(u)=e^u-1$. This leads to the  third equation of Theorem 1 which  is  yet another integrable deformation of the BF equation: its equivalent forms, $u_{xy}=(e^u)_{tt}-c(e^u)_{xx}$ and $u_{xy}=(e^u)_{tt}+(e^u)_{ty}$, have appeared in \cite{Fer, Dryuma}. Both can be reduced to case 3 via appropriate  (complex) linear transformations of the independent variables $x, y, t$. An alternative first order form of this equation  is provided by the system $v_t=\frac{v_y+w_y}{v+w}, \ w_x=\frac{v_y+w_y}{v+w}$ which appeared in \cite {FMS} in the classification of integrable Hamiltonian systems of hydrodynamic type in 2+1 dimensions. Setting $v=s_{xy}, w=s_{ty}$ we obtain the second order PDE $s_{xy}+s_{ty}=e^{s_{xt}}$. For $u= s_{xt}$ this gives the equation $u_{xy}+u_{ty}=(e^u)_{xt}$ which is also equivalent to case 3.

\medskip

\noindent {\bf Case 4.}
$$
A=\left(\begin{array}{ccc}
0&1&0\\
1&0&0\\
0&0&1
\end{array}
\right), ~~~ 
B=\left(\begin{array}{ccc}
1&0&0\\
0&0&0\\
0&0&0
\end{array}
\right).
$$
The substitution of $V=A+f(u)B$ into the integrability condition (\ref{b}) leads to $k=0$ and $f''=0$, so that without any loss of generality one can take $f(u)=u$. This leads to the equation
$(uu_x)_x+2u_{xy}+u_{tt}=0$, which is equivalent to the dKP equation.

\medskip

\noindent {\bf Case 5.}
$$
A=\left(\begin{array}{ccc}
0&0&1\\
0&1&0\\
1&0&0
\end{array}
\right), ~~~ 
B=\left(\begin{array}{ccc}
0&1&0\\
1&0&0\\
0&0&0
\end{array}
\right).
$$
The substitution of $V=A+f(u)B$ into the integrability condition (\ref{b}) leads to $k=0$ and $f''=0$, so that without any loss of generality one can take $f(u)=u$. This leads to the last equation in Theorem 1. Setting $u=v_{xy}$ we obtain the equation $v_{xy}^2+v_{yy}+2v_{xt}=0$. In this form, it appeared in \cite{MaksEgor} as the simplest case in the classification of  integrable  hydrodynamic chains satisfying the so-called `Egorov' property.

This finishes the proof of Theorem 1.

\section{Reduction to Painlev\'e equations}

In this section we investigate systems (\ref{N}),
$$
gN'=V,  ~~~ g^2 \det N=\xi=const,
$$
 for all  canonical forms of Theorem 1, and reduce them to Painlev\'e transcendents. For the BF and dKP equations this was done in   \cite{Tod, Dunajski}: we include these results for the sake of completeness. We find it more convenient to work with  $\sigma$-forms of Painlev\'e equations 
as listed in \cite{JM}, Appendix C.

\medskip

\noindent {\bf Case 1: $u_{xx}+u_{yy}-\ln(1-e^u)_{yy}-\ln(1-e^u)_{tt}=0.$} We have
$$
V=\left(\begin{array}{ccc}
1&0&0\\
0&\frac{1}{1-e^u}&0\\
0&0&\frac{e^u}{1-e^u}
\end{array}
\right), ~~~ 
N=\left(\begin{array}{ccc}
F&\alpha&\beta\\
\alpha&G&\gamma\\
\beta &\gamma&H
\end{array}
\right),
$$
so that the first equation (\ref{N}) implies that   $\alpha, \beta, \gamma$ are constants, while $F'=\frac{1}{g}, \ G'=\frac{1}{(1-e^u)g}, \ H'=\frac{e^u}{(1-e^u)g}$. Thus, $G'=\frac{1}{1-e^u}F', \ H'=\frac{e^u}{1-e^u}F'$. Setting $s=\frac{1}{1-e^u}$ one obtains $G_s=sF_s, \ H_s=(s-1)F_s$.
Parametrizing these relations in the form
$F=\sigma_s, \ G=s\sigma_s-\sigma, \ H=s\sigma_s-\sigma+\mu-\sigma_s$,  and substituting into the second equation (\ref{N}),
$$
FGN-\alpha^2H-\beta^2G-\gamma^2F+2\alpha \beta \gamma=\xi F'^2,
$$
 (note that $F'=s(s-1)F_s$), we obtain an ODE for $\sigma(s)$, 
$$
\xi s^2(s-1)^2\sigma_{ss}^2=\sigma_s(s\sigma_s-\sigma)(s\sigma_s-\sigma+\mu-\sigma_s)-\gamma^2\sigma_s-\beta^2(s\sigma_s-\sigma)-\alpha^2(s\sigma_s-\sigma+\mu-\sigma_s)+2\alpha \beta \gamma.
$$
This ODE belongs to the class
$$
s^2(s-1)^2\sigma_{ss}^2=a\sigma_s (s\sigma_s-\sigma)^2+(s\sigma_s-\sigma)P_2(\sigma_s)+P_1(\sigma_s),
$$
where $P_2$  and $P_1$ are  polynomials in $\sigma_s$ of the order two and one, respectively, such that the leading term of $P_2$ is $-a\sigma_s^2$. Modulo transformations $\sigma\to a_1\sigma+a_2$ this form is equivalent to the general case of $P_{VI}$,
$$
\sigma_s s^2(s-1)^2\sigma_{ss}^2+2[\sigma_s (s\sigma_s-\sigma)-\sigma_s^2-\nu_1\nu_2\nu_3\nu_4]^2=(\sigma_s+\nu_1^2)(\sigma_s+\nu_2^2)(\sigma_s+\nu_3^2)(\sigma_s+\nu_4^2).
$$
see \cite{JM}.

\medskip

\noindent {\bf Case 2: $u_{xx}+u_{yy}+(e^u)_{tt}=0.$}  We have
$$
V=\left(\begin{array}{ccc}
1&0&0\\
0&1&0\\
0&0&e^u
\end{array}
\right), ~~~ 
N=\left(\begin{array}{ccc}
F&\alpha&\beta\\
\alpha&G&\gamma\\
\beta &\gamma&H
\end{array}
\right),
$$
so that the first equation (\ref{N}) implies that   $\alpha, \beta, \gamma$ are constants, while $F'=1/g, \ G'=1/g, \ H'=e^u/g$. Thus, $ G'=F', \ H'=e^uF'$. Setting $e^u=s$, parametrizing these relations in the form
$F=\sigma_s, \ G=\sigma_s+\mu, \ H=s\sigma_s-\sigma$,  and substituting into the second equation (\ref{N}) we obtain an ODE for $\sigma(s)$, 
$$
\xi s^2\sigma_{ss}^2=\sigma_s(\sigma_s+\mu) (s\sigma_s-\sigma)-\gamma^2\sigma_s-\alpha^2(s\sigma_s-\sigma)-\beta^2(\sigma_s+\mu)+2\alpha \beta \gamma.
$$
This ODE belongs to the class
$$
s^2\sigma_{ss}^2= (s\sigma_s-\sigma)P_2(\sigma_s)+P_1(\sigma_s),
$$
where $P_2$  and $P_1$ are  polynomials in $\sigma_s$ of the order two and one, respectively. This is the special  case of $P_{V}$ which is reducible to $P_{III}$, see Case 3 below.

\medskip

\noindent {\bf Case 3: $(e^u-u)_{xx}+2u_{xy}+(e^u)_{tt}=0.$} We have
$$
V=\left(\begin{array}{ccc}
e^u-1&1&0\\
1&0&0\\
0&0&e^u
\end{array}
\right), ~~~ 
N=\left(\begin{array}{ccc}
F&G&\alpha\\
G&\beta&\gamma\\
\alpha &\gamma&H
\end{array}
\right),
$$
so that the first equation (\ref{N}) implies that   $\alpha, \beta, \gamma$ are constants, while $F'=(e^u-1)/g, \ G'=1/g, \ H'=e^u/g$. Thus, $H'=e^uG', \ F'=(e^u-1)G'$. Setting $e^u=s$, parametrizing these relations in the form
$G=\sigma_s, \ H=s \sigma_s-\sigma, \ F=s\sigma_s-\sigma+\mu-\sigma_s$,  and substituting into the second equation (\ref{N}) we obtain an ODE for $\sigma(s)$, 
$$
\xi s^2\sigma_{ss}^2=\beta (s\sigma_s-\sigma+\mu-\sigma_s)(s\sigma_s-\sigma)-\gamma^2(s\sigma_s-\sigma+\mu-\sigma_s)-\sigma_s^2(s\sigma_s-\sigma)+2\alpha \gamma \sigma_s-\alpha^2\beta.
$$
This ODE belongs to the class
\begin{equation}
s^2\sigma_{ss}^2= a(s\sigma_s-\sigma)^2+(s\sigma_s-\sigma)P_2(\sigma_s)+P_1(\sigma_s),
\label{P5}
\end{equation}
where $P_2$  and $P_1$ are  polynomials in $\sigma_s$ of the order two and one, respectively. In the case $a\ne 0$,  transformations  $\sigma \to a_1\sigma+a_2s+a_3, \ s \to b_1s$ bring any generic equation of the form (\ref{P5}) to the  canonical $P_{V}$ form,
$$
s^2\sigma_{ss}^2=[\sigma-s\sigma_s+2\sigma_s^2+(\nu_1+\nu_2+\nu_3)\sigma_s]^2 -4\sigma_s(\nu_1+\sigma_s)(\nu_2+\sigma_s)(\nu_3+\sigma_s).
$$
The case $a=0$ leads to the special  case of $P_{V}$ which is reducible to $P_{III}$,
$$
s^2\sigma_{ss}^2= (s\sigma_s-\sigma)P_2(\sigma_s)+P_1(\sigma_s).
$$

\medskip

\noindent {\bf Case 4: $u_{xt}-(uu_x)_x-u_{yy}=0.$} We have
$$
V=\left(\begin{array}{ccc}
-u&0&1/2\\
0&-1&0\\
1/2&0&0
\end{array}
\right), ~~~ 
N=\left(\begin{array}{ccc}
F&\alpha&G\\
\alpha&H&\beta\\
G&\beta&\gamma
\end{array}
\right),
$$
so that the first equation (\ref{N}) implies that   $\alpha, \beta, \gamma$ are constants, while $F'=-u/g, \ G'=1/(2g), \ H'=-1/g$. Thus, $G'=-H'/2, \ F'=uH'$. Parametrizing these relations in the form
$H=\sigma', \ H=\mu-\sigma'/2, \ F=u\sigma'-\sigma$, and substituting into the second equation (\ref{N}) we obtain an ODE for $\sigma(u)$, 
$$
\xi(\sigma'')^2=\gamma \sigma'(u\sigma'-\sigma)-\beta^2(u\sigma'-\sigma)- \sigma'(\mu-\sigma'/2)^2+2\alpha \beta (\mu-\sigma'/2)-\alpha^2\gamma.
$$
This ODE belongs to the class
\begin{equation}
(\sigma'')^2= a\sigma'(u\sigma'-\sigma)+b(u\sigma'-\sigma)+P_3(\sigma'),
\label{P2}
\end{equation}
where $P_3$ is a third order polynomial in $\sigma'$. In the case $a\ne 0$ one can eliminate $b$ by a transformation  $\sigma\to \sigma +s u$. Further transformations $\sigma \to a_1\sigma+a_2, \ u \to b_1u+b_2$ bring any equation of the form (\ref{P2}) to the canonical  $P_{II}$ form,
$$
(\sigma'')^2+ 4(\sigma')^3+2\sigma'(u\sigma'-\sigma)+\delta=0.
$$
Similarly, for $a=0,\ b\ne 0$, transformations $\sigma \to a_1\sigma +su+a_2, \ u \to b_1u+b_2$ bring any equation of the form (\ref{P2}) to the canonical $P_{I}$ form, 
$$
(\sigma'')^2+ 4(\sigma')^3+2(u\sigma'-\sigma)=0.
$$

\medskip

\noindent {\bf Case 5: $(u^2)_{xy}+u_{yy}+2u_{xt}=0.$} We have
$$
V=\left(\begin{array}{ccc}
0&u&1\\
u&1&0\\
1&0&0
\end{array}
\right), ~~~ 
N=\left(\begin{array}{ccc}
\alpha&F&G\\
F&H&\beta\\
G&\beta&\gamma
\end{array}
\right),
$$
so that the first equation (\ref{N}) implies that   $\alpha, \beta, \gamma$ are constants, while $G'=1/g, \ H'=1/g, \ F'=u/g$. Thus, $H'=G', \ F'=uG'$. Parametrizing these relations in the form
$G=\sigma', \ H=\sigma'+\mu, \ F=u\sigma'-\sigma$, and substituting into the second equation (\ref{N}) we obtain an ODE for $\sigma(u)$, 
$$
\xi(\sigma'')^2=-\gamma (u\sigma'-\sigma)^2+2\beta \sigma'(u\sigma'-\sigma)-\sigma'^2(\sigma'+\mu)+\alpha \gamma (\sigma'+\mu)-\alpha\beta^2.
$$
This ODE belongs to the class
\begin{equation}
(\sigma'')^2= a(u\sigma'-\sigma)^2+b\sigma'(u\sigma'-\sigma)+P_3(\sigma'),
\label{P4}
\end{equation}
where $P_3$ is a third order polynomial in $\sigma'$. In the case $a\ne 0$ one can eliminate $b$ by a translation of $u$. Further transformations $\sigma \to a_1\sigma+a_2u, \ u \to b_1u$ bring any equation of the form (\ref{P4}) to the canonical  $P_{IV}$ form,
$$
(\sigma'')^2= 4(u\sigma'-\sigma)^2-4\sigma'(\sigma'+\nu_1)(\sigma'+\nu_2),
$$
Similarly, for $a=0,\ b\ne 0$, transformations $\sigma \to a_1\sigma +a_3, \ u \to b_1u+b_2$ bring any equation of the form (\ref{P4}) to the canonical $P_{II}$ form, 
$$
(\sigma'')^2+ 4(\sigma')^3+2\sigma'(u\sigma'-\sigma)+\delta=0.
$$

\section{Central quadric ansatz versus hydrodynamic reductions}

The main observation of this section is that solutions coming from the central quadric ansatz (\ref{central}) can also be obtained as a particular case of two-phase solutions,
$$
u=u(R^1, R^2),
$$
where the phases $R^1(x, y, t), \ R^2(x, y, t)$ satisfy a pair of two-component systems of hydrodynamic type,
\begin{equation}
R^i_t = \lambda^i(R) R^i_x, \quad R^i_y = \mu^i(R) R^i_x,
\label{2}
\end{equation}
$i=1, 2$.  The general solution of equations (\ref{2}) is given by the so-called  generalised hodograph formula \cite{Tsarev},
\begin{equation}
\begin{array}{c}
x+\lambda^1 (R^1, R^2) t+ \mu^1 (R^1, R^2) y=\nu^1 (R^1, R^2),\\
x+\lambda^2 (R^1, R^2) t+ \mu^2 (R^1, R^2) y=\nu^2 (R^1, R^2),\\
\end{array}
\label{hod}
\end{equation}
where $\nu^i$ are the characteristic speeds of yet another flow, 
\begin{equation}
R^i_{\tau}=\nu^i(R)R^i_x,
\label{nu}
\end{equation}
which commutes with (\ref{2}). 
Equations (\ref{hod}) define $R^1(x, y, t)$ and $R^2(x, y, t)$ implicitly. These relations can also be viewed as  the equations of a line congruence (two-parameter family of lines) 
in the space of independent variables $x, y, t$, parametrised by $R^1, R^2$. As $u=u(R^1, R^2)$ is constant along the lines of this congruence, the level surfaces $u=const$ will automatically be ruled. Since a quadric carries two (complex) one-parameter families of lines, solutions constant on quadrics can therefore be viewed as two-phase solutions in `two different ways': they are constant along  two distinct congruences of lines.

Taking dKP equation as an example, we will now provide explicit derivation of  the central quadric ansatz  from the general two-phase solutions. Let us rewrite  dKP  as a two-component first order system,
$$
u_t=uu_x+w_y, ~~~ u_y=w_x.
$$
It will be more convenient for our purposes to work with variables $u, w$ rather than $R^1, R^2$. 

\medskip

\noindent {\bf Proposition 1.} {\it Two-phase solutions of the dKP equation, $u(x, y, t)$ and  $w(x, y, t)$, are given by the implicit formulae
\begin{equation}
x+(z_u+u)t=m_u, ~~~ y+z_w t=m_w,
\label{hod1}
\end{equation}
where the functions $z(u, w)$ and $m(u, w)$  satisfy the PDEs
\begin{equation}
\begin{array}{c}
z_{uu}+z_wz_{uw}-z_uz_{ww}+1=0,\\
m_{uu}+z_wm_{uw}-z_um_{ww}=0.
\end{array}
\label{zm}
\end{equation}
}

\centerline{\bf Proof:}

\medskip

The analogues of equations (\ref{2})  read
\begin{equation}
u_t = uu_x+z_x,  \ w_t=q_x ~~~~~~ {\rm and}    ~~~~~~~ u_y=w_x, \  w_y=z_x, 
\label{ty}
\end{equation}
where the commutativity conditions, $u_{ty}=u_{yt}$ and  $w_{ty}=w_{yt}$, imply that the functions $q(u, w)$ and $z(u, w)$  satisfy the relations $q_u=z_uz_w, \ q_w=z_w^2+z_u+u$. Notice that the compatibility condition, $q_{uw}=q_{wu}$, implies the first relation (\ref{zm}). This equation  for $z(u, w)$ first appeared in \cite{GibTsa96} in the context of hydrodynamic reductions of Benney's moment equations. It is known as the Monge-Amp\`ere form of the Gibbons-Tsarev system. The analogue of the third commuting flow (\ref{nu}) is
\begin{equation}
u_{\tau}=m_x, \ w_{\tau}=n_x,
\label{tau}
\end{equation}
where the commutativity conditions imply that  the functions $m(u, w)$ and $n(u, w)$  satisfy the relations $n_u=z_um_w, \ n_w=z_wm_w+m_u$. Their compatibility condition, $n_{uw}=n_{wu}$, implies the second relation (\ref{zm}). Rewriting equations (\ref{ty}) and (\ref{tau}) in matrix form we obtain
$$
\left(\begin{array}{c}
u\\
w
\end{array}
\right)_t=
\left(\begin{array}{cc}
z_u+u & z_w\\
z_uz_w & z_w^2+z_u+u
\end{array}
\right)
\left(\begin{array}{c}
u\\
w
\end{array}
\right)_x, 
~~~~~ 
\left(\begin{array}{c}
u\\
w
\end{array}
\right)_y=
\left(\begin{array}{cc}
0 & 1\\
z_u & z_w
\end{array}
\right)
\left(\begin{array}{c}
u\\
w
\end{array}
\right)_x, 
$$
and 
$$
\left(\begin{array}{c}
u\\
w
\end{array}
\right)_{\tau}=
\left(\begin{array}{cc}
m_u & m_w\\
z_um_w & z_wm_w+m_u
\end{array}
\right)
\left(\begin{array}{c}
u\\
w
\end{array}
\right)_x, 
$$
respectively. In matrix notation, the generalised hodograph formula (\ref{hod}) is
$$
\left(\begin{array}{cc}
1 & 0\\
0 & 1
\end{array}
\right)x+
\left(\begin{array}{cc}
z_u+u & z_w\\
z_uz_w & z_w^2+z_u+u
\end{array}
\right)t+
\left(\begin{array}{cc}
0 & 1\\
z_u & z_w
\end{array}
\right)y
=\left(\begin{array}{cc}
m_u & m_w\\
z_um_w & z_wm_w+m_u
\end{array}
\right).
$$
One can readily see that only two of these relations are independent, and they are equivalent to (\ref{hod1}). 

\medskip

\noindent{\bf Remark.}  Differentiating   (\ref{hod1}) implicitly by $x, y, t$ we obtain the relations
\begin{equation}
\label{eq:diff-invariance}
u_t=(z_u+u)u_x+z_wu_y, ~~~ w_t=(z_u+u)w_x+z_ww_y,
\end{equation}
which manifest the invariance of the corresponding solutions under the conditional symmetry \begin{equation}
X=   (z_u+u)\partial_x +z_w\partial_y-\partial_t. 
\end{equation}
The presence of such symmetry is characteristic of two-phase solutions for equations in $2+1$ dimensions  \cite{Grundland-Huard-2007,Huard-2010}. 
Differentiating (\ref{eq:diff-invariance}) with respect to $x,y, t$ and eliminating  the derivatives of $z$ we obtain second order differential constraints  for $u$ and $w$ characterising two-phase solutions,
\begin{equation}
\label{eq:constraints-2nd-order}
\begin{split}
\delta_{xy}^2 u_{tt} + \delta_{xt}^2 u_{yy} + \delta_{yt}^2 u_{xx} - 2 \left( \delta_{xt} \delta_{xy} u_{ty} + \delta_{tx} \delta_{ty} u_{xy} + \delta_{yt} \delta_{yx} u_{tx}\right) = 0,\\
\delta_{xy}^2 w_{tt} + \delta_{xt}^2 w_{yy} + \delta_{yt}^2 w_{xx} - 2 \left( \delta_{xt} \delta_{xy} w_{ty} + \delta_{tx} \delta_{ty} w_{xy} + \delta_{yt} \delta_{yx} w_{tx}\right) = 0,\\
\end{split}
\end{equation}
where the functions $\delta_{yt},\delta_{tx},\delta_{xy}$ are $2\times 2$ minors of the Jacobi matrix $J=\partial (u,w)/\partial(x,y,t)$, i.e. $\delta_{yt}=u_yw_t-u_tw_y$, etc. Introducing the vector field $X=(\delta _{yt}, \delta_{tx}, \delta_{xy})$ one can rewrite these constraints  in the equivalent form
$XUX^T=0, \ XWX^T=0$ where $U, W$ are the Hessian matrices of $u$ and $w$.
The constraints (\ref{eq:constraints-2nd-order}) completely characterize two-phase solutions of the dKP equation, providing an easy criterion to verify if a given solution can be constructed using this method. We emphasize that  differential constraints (\ref{eq:constraints-2nd-order}) are {\it universal}: they govern two-phase solutions of {\it any} two-component quasilinear system. Indeed, the general solution of (\ref{eq:constraints-2nd-order}) is given by implicit relations
\begin{equation}
 x+p(u, w)t=q(u, w), ~~~ y+r(u, w) t=s(u, w),
\label{imp}
\end{equation}
where $p, q, r, s$ are arbitrary functions. In other words,  functions   $u, w$ solve (\ref{eq:constraints-2nd-order}) if and only if they are constant along a two-parameter family of lines. This follows from the following equivalent geometric representation of the constraints  (\ref{eq:constraints-2nd-order}):
$$
L_Xu=L_Xw=0, ~~~ XUX^T=XWX^T=0.
$$
The first two conditions mean that the functions $u$ and $w$ are constant along integral trajectories of the vector field $X$. This, in particular,  implies that $X=(\delta _{yt}, \delta_{tx}, \delta_{xy})$.
The last two conditions, which are equivalent to (\ref{eq:constraints-2nd-order}), mean that integral trajectories of $X$ are asymptotic curves on the level surfaces $u=const$ and $w=const$. 
It remains to point out that if two surfaces intersect transversally along a curve which is an asymptotic curve on both of them,  this curve must be a straight line (indeed, its osculating plane is tangential to both surfaces, and hence  degenerates into a line).
Adding the relations (\ref{imp}) to the dKP equation one obtains constraints for $p, q, r, s$ resulting in (\ref{hod1}).


\medskip

Following \cite{Kaptsov} we will demonstrate that the system (\ref{zm}) possesses solutions expressible in terms of the Painlev\'e equation $P_{II}$.

\medskip

\noindent {\bf Proposition 2.} {\it The system (\ref{zm}) possesses solutions of the form
$$
z(u, w)=s_1(u)+s_2(u)e^w+s_3(u)e^{-w}, ~~~ m(u, w)=s_2(u)e^w-s_3(u)e^{-w},
$$
where the functions $s_1(u), s_2(u), s_3(u)$ satisfy the ODEs
\begin{equation}
s_1'=2s_2s_3-u, ~~~
s_2''=(2s_2s_3-u)s_2, ~~~
s_3''=(2s_2s_3-u)s_3.
\label{s}
\end{equation}
}
This expression for $z$ was obtained in \cite{Kaptsov} based on the method of linear differential constraints. Substituting the above  $z, m$ into the equations (\ref{ty}) and eliminating $w$, one obtains an implicit relation of the form (\ref{central}) for $u$,
\begin{equation}
(s_2s_3)'^2(t^2-1)-s_2s_3(x+2s_2s_3t)^2+(s_3s_2'-s_2s_3')(x+2s_2s_3t)y+s_2's_3'y^2=0,
\label{s1}
\end{equation}
which shows that $u$ is indeed constant on central quadrics. It remains to demonstrate that the system (\ref{s}) is equivalent to $P_{II}$. Indeed, equations (\ref{s}) possess a conservation law, $s_3s_2'-s_2s_3'=c=const$, which can be resolved in the form $s_2'=ps_2+c/s_3, \ s_3'=ps_3$. This, in particular, implies the relation $(s_2s_3)'=2ps_2s_3+c$. Differentiating $s_3'=ps_3$ and using  the last equation (\ref{s}) we obtain $2s_2s_3=p'+p^2+u$. Substituting this expression for $s_2s_3$ into the previous relation we obtain the $P_{II}$ equation,
$$
p''=2p^3+2pu+2c-1.
$$
Ultimately, all coefficients of the relation (\ref{s1}) are expressed in terms of $p$,
$$
s_2s_3=\frac{1}{2}(p'+p^2+u), ~~~ s_3s_2'-s_2s_3'=c, ~~~ s_2's_3'=p^2s_2s_3+cp=\frac{1}{2}p^2(p'+p^2+u)+cp.
$$
This gives  a solution equivalent to the one constructed in \cite{Dunajski}.

\section*{Acknowledgements}

We thank A. Bolsinov,   M. Mazzocco,  M. Pavlov  and A. Veselov for clarifying discussions. The research of EVF was partially supported  by the European Research Council Advanced Grant  FroM-PDE. The research of BH was supported by a FQRNT  scholarship.

\end{document}